\documentclass[aps,prl,twocolumn,amsmath,amssymb,nofootinbib,superscriptaddress]{revtex4-2}

\usepackage{times}
\usepackage[pdftex]{graphicx}
\usepackage{dcolumn}
\usepackage{bm}
\usepackage{amsmath}
\usepackage{indentfirst}
\usepackage{float}
\usepackage[colorlinks]{hyperref}
\usepackage[dvipsnames]{xcolor}
\usepackage{xcolor}
\usepackage{subfigure}
\usepackage{algorithm}
\usepackage{algorithmicx}
\usepackage{algpseudocode}
\usepackage{amsmath}
\usepackage{verbatim}
\usepackage{makecell}
\usepackage[normalem]{ulem}

\usepackage{accents}
\usepackage{braket}
\usepackage{booktabs}  
\usepackage{threeparttable}

\newcommand{\trace}{{\rm tr}}
\newcommand{\voca}{\mathcal{V}}
\newcommand{\rhoop}{\rho}
\newcommand{\sos}{\left <sos \right >}
\newcommand{\eos}{\left <eos \right >}
\newcommand{\pad}{\left <pad \right >}
\newcommand{\Hop}{H}
\newcommand{\fidelity}{\mathcal{F}}
\newcommand{\loss}{\mathcal{L}}
\newcommand{\Nmax}{N_{\max}}
\newcommand{\samples}{\mathcal{S}}

\def\sos{\operatorname{sos}}
\def\eos{\operatorname{eos}}
\def\pad{\operatorname{pad}}
\def\true{\operatorname{true}}
\def\s{\operatorname{s}}

\begin{document}

\title{Quantum State Tomography Inspired by Language Modeling}% Force line breaks with \\

\author{Lu Zhong}

%  \homepage{http://www.Second.institution.edu/~Charlie.Author}
\affiliation{
 Institute of Fundamental and Frontier Sciences,\\
 University of Electronic Science and Technology of China, Chengdu, 610051, China% with \\
}%

\author{Chu Guo}
%  \altaffiliation[Also at ]{Physics Department, XYZ University.}%Lines break automatically or can be forced with \\
\email{guochu604b@gmail.com}
% \affiliation{Henan Key Laboratory of Quantum Information and Cryptography, Zhengzhou,
% Henan 450000, China}
\affiliation{%
 Key Laboratory of Low-Dimensional Quantum Structures and Quantum Control of Ministry of Education,\\
 Department of Physics and Synergetic Innovation Center for Quantum Effects and Applications,\\
 Hunan Normal University, Changsha 410081, China\\
}%

\author{Xiaoting Wang}
\email{xiaoting@uestc.edu.cn}
\affiliation{
 Institute of Fundamental and Frontier Sciences,\\
 University of Electronic Science and Technology of China, Chengdu, 610051, China% with \\
}%

% \author{Lu Zhong}
% \email{202021210215@std.uestc.edu.cn}
% \affiliation{%
% Institute of Fundamental and Frontier Sciences,\\
%  University of Electronic Science and Technology of China, Chengdu, 610051, China
% }%

%              %  but any date may be explicitly specified

\begin{abstract}
Quantum state tomography is an elementary tool to fully characterize an unknown quantum state. As the quantum hardware scales up in size, the standard quantum state tomography becomes increasingly challenging due to its exponentially growing complexity. In this work, we propose a scalable solution by considering state tomography as a language modeling task, where the unknown quantum state is treated as an unknown language, the correlation of the quantum state is interpreted as the semantic information specific to this language, and the measurement outcomes are simply the text instances generated from the language. Based on a customized transformer model from language modeling, we demonstrate that our method can accurately reconstruct prototypical pure and mixed quantum states using less samples than state-of-the-art methods. More importantly, our method can reconstruct a class of similar states simultaneously, in comparison with the existing neural network methods that need to train a model for each unknown state.
\end{abstract}

\maketitle

\textit{Introduction} -- Quantum state tomography (QST) reconstructs an unknown quantum state by measuring identically prepared copies, a fundamental ingredient for quantum information processing. Standard QST works by measuring the unknown quantum state in an informationally complete basis set and fitting the best density matrix using maximum likelihood estimation~\cite{WhiteKwiat1999,JamesWhite2001}. Due to its exponentially growing complexity in both the number of quantum measurements and the classical post-processing, the reachable system size by standard QST is limited within a few qubits~\cite{LuPan2007,Haffner2005scalable}. In the meantime, quantum computing technologies have been developing rapidly in recent years. As a milestone, quantum computational advantages with up to $60$ qubits have already been demonstrated for the task of random quantum circuit sampling~\cite{AruteMartinis2019,WuPan2021,ZhuPan2022}. The giant gap between the experimentally controllable number of qubits and the feasible number of qubits for which one can perform high-fidelity QST calls for scalable QST methods.

In general, any QST scheme which is more efficient than the standard QST has to make certain assumptions on the unknown quantum state~\cite{TothWeinfurter2010,MoroderWeinfurter2012,GrossEisert2010,LiuYuan2012,SmithJessen2013,RiofrioEisert2017} or only aims to extract partial information of the quantum state instead of the whole state itself~\cite{Aaronson2018,HuangJohn2020,ChenSteven2021,GuptaKais2021}. Such examples include permutationally invariant QST~\cite{TothWeinfurter2010,MoroderWeinfurter2012} and compressed sensing QST~\cite{GrossEisert2010,LiuYuan2012,SmithJessen2013,RiofrioEisert2017}, which take advantage of the symmetry or the sparsity of the unknown quantum state. For (fairly) pure and slightly entangled quantum states such that they can be efficiently represented as Matrix Product States (MPS), there exist polynomial reconstruction algorithms with guaranteed convergence~\cite{CramerLiu2010,BaumgratzPlenio2013,LanyonRoos2017,Guo2022a}. However, the generalization of this approach to mixed quantum states~\cite{TaoRaymond2017} or higher-dimensional setups is not straightforward. In contrast, state-of-the-art quantum processors all possess a two-dimensional structure, and can easily generate highly entangled (but noisy) quantum states~\cite{AruteMartinis2019,WuPan2021,ZhuPan2022}. 

Recently the embracement of neural network methods and QST has produced very fruitful results~\cite{TorlaiCarleo2018,TorlaiMelko2018,XinZeng2019,PalmieriKulik2020,GrayBose2018,ShangGuo2019,MelkaniNori2020,WeissIsart2019,TaoJun2020,LiuWu2020,XueWu2022}. In these approaches, the unknown quantum state is represented by a neural network ansatz whose parameters are then variationally optimized~\cite{CarleoTroyer2017}. Neural network quantum states are more expressive than MPS~\cite{SharirCarleo2021} and can efficiently represent certain volume-law entangled quantum states~\cite{DengSarma2017}. Neural network-based QST has made the least stringent assumptions on the underlying quantum states while showing tremendous potential as a practical and scalable QST scheme. Concretely, neural network-based QST methods have been applied to both pure quantum states~\cite{TorlaiCarleo2018} and mixed quantum states~\cite{TorlaiMelko2018,CarrasquillaAolita2019}, using a variety of neural network ansatz ranging from the restricted Boltzmann machine~\cite{TorlaiCarleo2018,TorlaiMelko2018,CarrasquillaAolita2019}, variational antoencoder~\cite{RocchettoSeverini2018}, to generative adversarial networks~\cite{AhmedKockum2021}. However given an unknown quantum state, it is generally unclear which neural network ansatz to use. Thus, it would be helpful to develop a neural network ansatz that has been proven to be universally applicable and successful for similar classical problems to QST.

In this work, we draw the similarity between QST and the language modeling task. In language modeling, one tries to make the model learn the syntax rules of an unknown language from a set of observed text instances. Once the unknown language is learned, the model could automatically ``speak'', namely generate new text instances~\cite{Chowdhary2020}. In QST, the unknown many-qubit quantum state behaves like an unknown language, for which the vocabulary consists of all the possible local measurement outcomes (labeled by integers). Upon measurement, one observes a specific sequence of integers. The quantum correlation between different qubits dictates that the integers at different positions of the sequence are generally correlated, such that only sequences following specific patterns can be observed. Once the language model is successfully trained for the unknown quantum state, it will be able to generate new integer sequences which share the same pattern as the observed ones used for training. In our QST proposal, we will design a customized neural network following this idea to solve the corresponding language modeling task. We name this method language-modeling quantum state tomography (LM-QST) for convenience. Our numerical simulations show that our method can efficiently reconstruct a family of unknown states using training data with a size no more than $O(10^4)$ for problems with up to 50 qubits.

\textit{LM-QST model} -- Before proceeding with our LM-QST method, it is notable that for neural network-based QST, one could either use a neural network ansatz to represent the unknown quantum state~\cite{TorlaiCarleo2018} directly or to represent the probability distribution of the measurement outcomes indirectly~\cite{CarrasquillaAolita2019}. In LM-QST, we will take the second approach, that is, for an $N$-qubit unknown quantum state denoted as $\rhoop$, we aim to reconstruct the $N$-variate probability distribution $P(\bold{a})$ defined as: 
\begin{align}\label{eq:Pdef}
P(\bold{a}) \equiv \trace(M^{\bold{a}} \rho),
\end{align}
Here, $\bold{a}$ denotes a specific measurement outcome written as $\bold{a} = (a_1, a_2, \dots, a_N)$, with $a_n$ an integer representing the measurement outcome of a local \textit{informationally complete} POVM (IC-POVM) on the $n$-th qubit. We have also used $M^{\bold{a}} = M^{a_1}\otimes M^{a_2} \otimes \cdots \otimes M^{a_N}$, where $M^{a_n}$ is a specific measurement operator on the $n$-th qubit. In general, for an $m$-outcome IC-POVM, each $a_n$ can choose $m$ different values, labeled as $\{0, 1, \dots, m-1\}$, we will use $m=4$ throughout this work but our approach is completely general for any value of $m$. Therefore, $P(\bold{a})$ locates in a space of size $4^N$, the same as $\rho$, and they can be transformed into each other since $M^{\bold{a}}$ is an invertible matrix in the $4^N$-dimensional linear space. This transformation is more effortless for tensor network-based ansatz~\cite{HanWang2022} than the neural network-based ansatz. The latter is generally exponentially difficult. Thus, there will be an exponential classical post-processing overhead to obtain $\rhoop$ if our neural network ansatz is designed for $P(\bold{a})$ (See Ref.~\cite{CarrasquillaAolita2019} or Supplementary for more details of the conversion process).
% Thus, $P(\bold{a})$ is a distribution defined on a sample space no larger than $4^N$. Once $P(\bold{a})$ is obtained from LM-QST, we can efficiently reconstruct $\rho$ from it through tensor network based ansatz\cite{HanWang2022}.
% This transformation is effortless for tensor network-based ansatz~\cite{HanWang2022}, while it is generally exponentially difficult for neural network-based ansatz. Thus, there will be an exponential classical post-processing overhead to obtain $\rhoop$ if our neural network ansatz is designed for $P(\bold{a})$

\begin{figure}[ht]
    \centering
    \includegraphics[width=0.9\columnwidth]{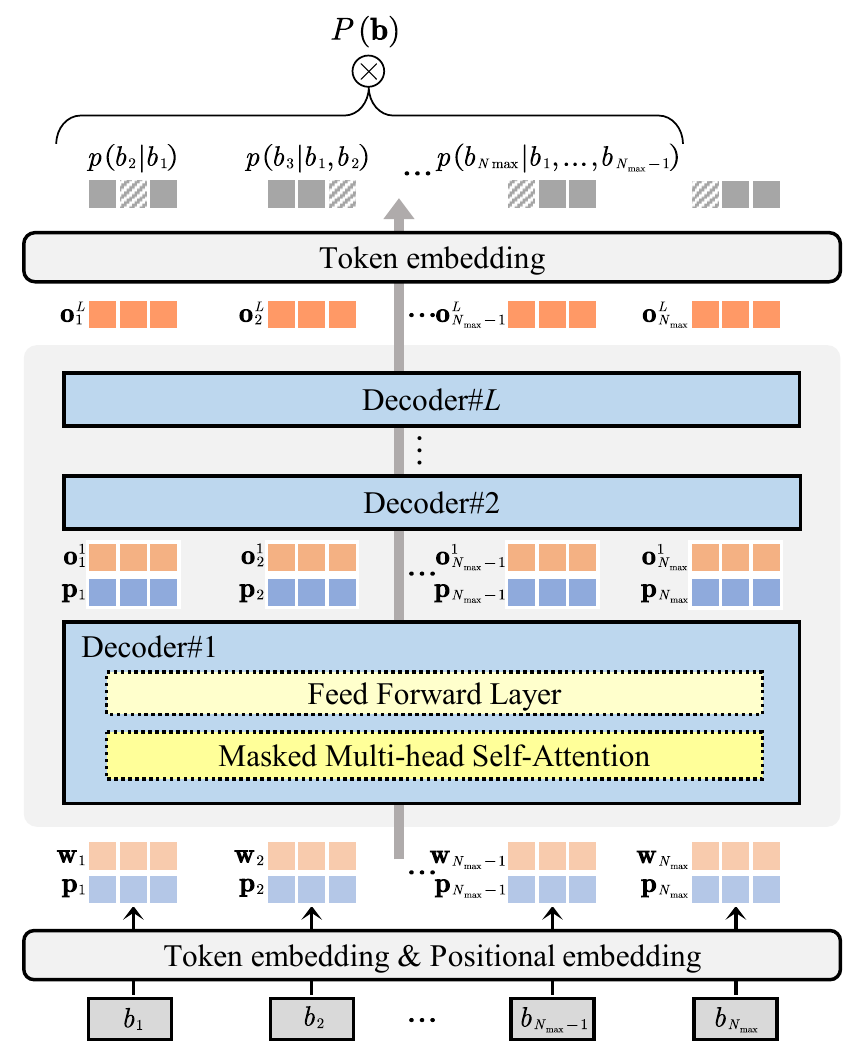}
    \caption{The macroscopic architecture of LM-QST. From bottom up, the model starts with an embedding layer which encodes each integer $b_j$ into a context vector $\bold{w}_j$ and a positional vector $\bold{p}_j$ with the embedding size $d$, which are then fed into $L$ sequentially stacked decoder blocks. The $l$-th decoder block accepts $\bold{p}_j$ and the context vectors $\bold{o}^{l-1}_j$ from the $l-1$-layer ($\bold{o}^0_j = \bold{w}_j$) as inputs and outputs a new sequence of context vectors $\bold{o}^l_j$. The output context vectors $o^L_j$ from the last decoder block are sent to a second embedding layer which outputs $\Nmax$ conditional probability distributions from $p(x_2|b_1)$ to $p(x_{\Nmax+1}|b_{1}, \dots, b_{\Nmax})$. $P(\bold{b})$ is obtained by selecting the elements $p(x_{j+1}=b_{j+1}|b_{1}, \dots, b_{j})$ for $1\leq j< \Nmax$ and multiply them together as in Eq.(\ref{eq:Pa}). The hyperparameters $L$, $\Nmax$ and $d$ fully determine the total number of tunable parameters in our LM-QST.
     }
\label{model-struct}
\end{figure}

Our proposed method begins with encoding measurement outcome sequences into datasets of expressions in an unknown language. In LM-QST method, denoting the set of the local IC-POVM outcomes as $\voca_0=\{0,1, \dots, m-1 \}$, we can construct the following vocabulary list $\voca$: 
\begin{align}\label{eq:vaco}
\voca \equiv \left \{0,1, \dots, m-1, \sos, \eos, \pad \right \}.
\end{align}
where different local measurement outcomes are encoded into distinctive tokens in $\voca$, with $|\voca|=m+3$. Specifically, any measurement outcome $\bold{a} =(a_1,\ldots,a_N)$, $a_k\in \voca_0$ of an $N$-qubit quantum system is encoded into a word sequence $\bold{b} =( \sos, a_1,\ldots,a_N, \eos)$, with the length of $|\bold{b}|=N+2$. In addition, $\pad$ token is the token used to force different sequences into the same length.

% \rem{We use our LM-QST to represent the multi-variate probability distribution}
The aim of LM-QST is to generate the probability distribution $P(\bold{b})$, where $\bold{b}= (b_1, b_2, \dots)$ is the encoded word sequence denoting the measurement outcome $\bold{a}$.  The detailed structure of the LM-QST network is shown in Fig.~\ref{model-struct}. First, the input $\bold{b}$ is fed into an embedding layer, with each $b_k$ embedded into a context vector $\bold{w}_k$ and a position vector $\bold{p}_k$, with an embedding size $d$. Then $\bold{w}_k$ and $\bold{p}_k$ are fed sequentially into $L$ stacked decoder blocks. Each decoder block takes all $\bold{p}_k$ and context vectors from the previous layer as inputs and outputs a new sequence of context vectors ($\bold{p}_k$ keeps unchanged). A final embedding layer takes the output context vectors of the final decoder block as inputs and outputs $\Nmax$ number of probability distributions, each with size $|\voca|$. The $j$-th output is interpreted as the conditional probability distribution of the $j+1$-th token on the previous tokens, denoted as $p(x_{j+1}|b_{1}, \dots, b_{j})$. Selecting each $x_j = b_j$, $P(\bold{b})$ can be computed as
\begin{align}\label{eq:Pa}
P(\bold{b}) = p(b_2|b_1) p(b_3|b_1,b_2)\dots p(b_{\Nmax}|b_1, b_2,\dots b_{\Nmax-1}),
\end{align}
where the last output $p(x_{\Nmax+1}|b_1, \dots, b_{\Nmax})$ is neglected (we usually have $b_1=\sos$ and $b_{\Nmax}=\eos$).
A model satisfying Eq.~(\ref{eq:Pa}) is said to be auto-regressive, and it allows for efficient and exact samplings~\cite{lipton2015critical}.
Next, based on the computed $P(\bold{b})$, the negative log-likelihood loss function is used for training, defined as
\begin{align}\label{eq:loss}
\loss = -\sum_{\bold{a} \in \samples} \ln P(\bold{a}),
\end{align}
where $\samples$ denotes the training dataset. Thus, the desired LM-QST model is derived after training.

Here, we stress that our model is designed as a GPT-2-inspired transformer variation for QST. The GPT-2~\cite{radford2019gpt2} evolves from the transformer architecture~\cite{2017attention}, which performs exceptionally in language modeling task among other neural network designs. The fundamental building block in our model is the customized attention-based decoder block as constructed in Fig.\ref{model-struct}, which consists of a masked multi-head self-attention layer followed by a feed-forward layer. The former is made to learn plausible rules underlying measurement outcome sequences. Essentially, it serves two purposes. First, before processing a given token, this layer can incorporate the model's understanding of related components by scoring the relevance between tokens and summing their vector representations, thus explaining the token's context information. Second, we developed a new training strategy for the positional vectors used in this layer. In order to emphasize the positional information between qubits and make it more appropriate for our QST task, we analyze the context and positional vectors independently(See Supplementary for details of the decoder blocks). Notice that attention-based models have achieved excellent success in recent years. Famous attention-based language models include BERT~\cite{devlin2018bert}, GPT series~\cite{radford2018gpt,radford2019gpt2}, and XLNet~\cite{yang2019xlnet}. Other than language modeling, attention-based architectures have also been used for computer vision~\cite{dosovitskiy2020vit, zhai2022lit}, biology~\cite{madaniProGenLanguageModeling2020, rivesBiologicalStructureFunction2021}, economics~\cite{liStockPricePrediction2018}. Since we have reduced QST as a language modeling task, we anticipate our LM-QST will function well for the QST task.
% It is indeed the case in the numerical simulations below.

\textit{LM-QST for single state} -- We first consider the standard single-state QST setting where the model is trained with $N_s$ fixed-length integer sequences from measuring the same quantum state. To evaluate the reconstruction accuracy, we further assume the actual value of a state $\rho$ is $\rho_{\true}$, for which we can calculate $P_{\true}(\bold{a})$ with Eq.~(\ref{eq:Pdef}). Then we calculate the classical fidelity between $P(\bold{a})$ and $P_{\true}(\bold{a})$:
\begin{align}\label{eq:fc}
\mathcal{F}_c(P,P_{\true}) \equiv \mathbb{E}_{\bold{a}\sim P}\left[\sqrt{\frac{P_{\true}(\bold{a})}{P(\bold{a})}}\right],
\end{align}
If $\mathcal{F}_c$ is very close to $1$, we can tell that $\rho$ is a state very close to $\rho_{\true}$. Alternatively, one could first reconstruct $\rho$ based on $P(\bold{a})$~\cite{CarrasquillaAolita2019}, and then calculate the quantum fidelity between $\rho$ and $\rho_{true}$, but this, in general, is computationally more expensive (more benchmarks of small systems using other measures including the quantum fidelity can also be found in Supplementary).
%This quantity can be efficiently computed by sampling the reconstructed $P(\bm{a})$, while evaluating the corresponding quantum fidelity is very difficult (it has been shown that heuristically the quantum fidelity has a qualitatively similar behavior as the classical fidelity~\cite{CarrasquillaAolita2019}, more benchmarks of small systems using other measures including the quantum fidelity can also be found in Supplementary).
%$\rho=\E(\rho') \equiv \left(1-\gamma\right)\rho + \frac{\gamma}{3}\left(  X\rho X + Y\rho Y + Z\rho Z\right )$.

\begin{figure}[ht]
    \centering
    \includegraphics[width=\columnwidth]{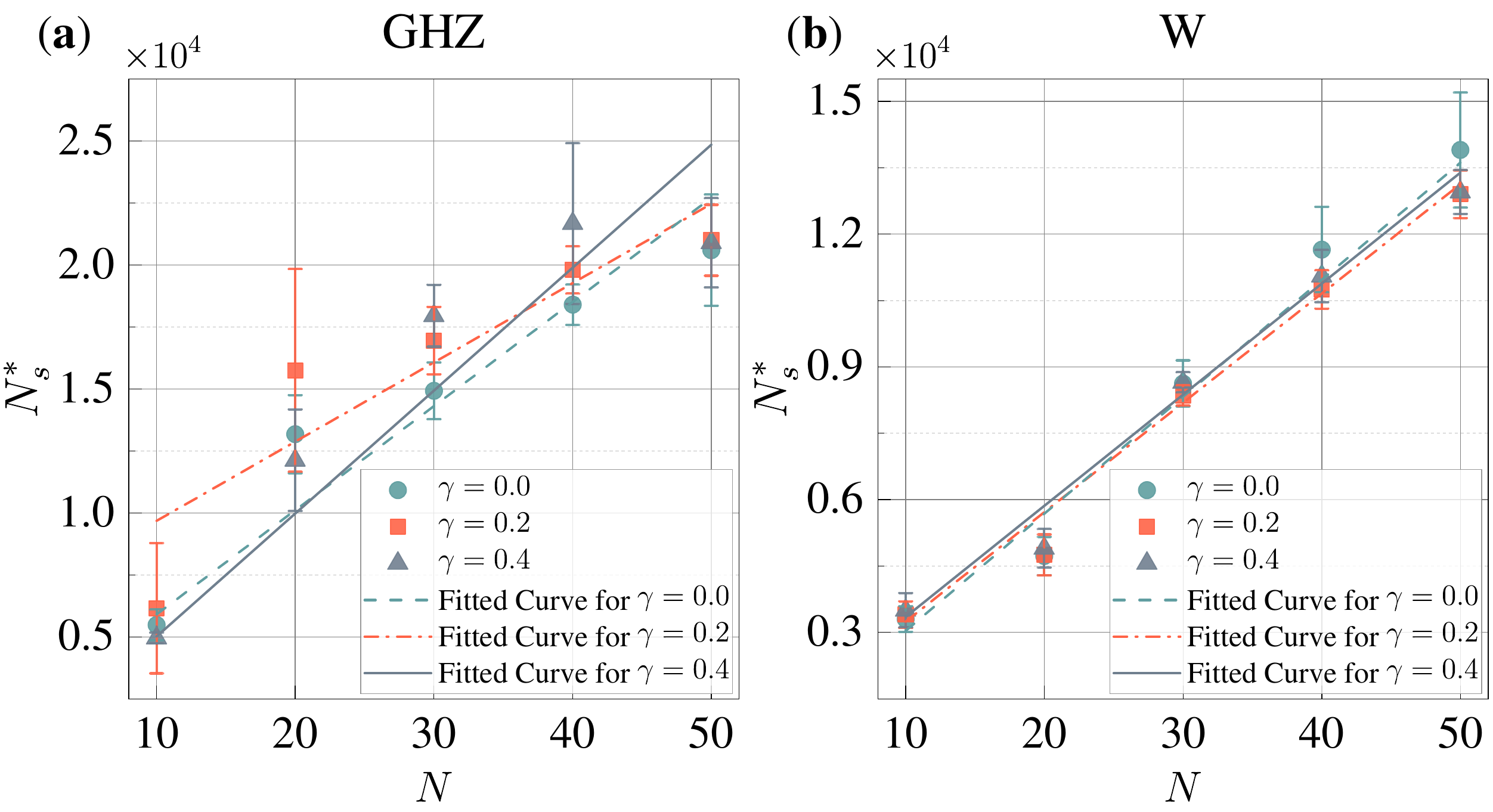}
    \caption{Scaling of the minimal number of samples $N_s^{\ast}$ required to reach $\fidelity_c \geq 99\%$ against the system size $N$ for (a) the GHZ states  and (b) the W states. The cadet-blue circles, tomato-red squares, and slate-gray triangles represent the mean values (error bars for the standard derivations) averaged over $20$ independently generated synthetic datasets for $\gamma=0,0.2,0.4$ respectively. The dashed, dot-dashed and solid lines with the same colors are the corresponding linear fittings. In these simulations we have used $L=4$ and $d=64$. 
    % \gcc{All the models have been trained with the same training settings with $d=64$, $L=4$, $h=4$. Learning rate have been set to $10^{-3}$, we also used Adam optimizer.}
    We have also used the Adam optimizer with an initial learn rate $10^{-3}$, and used $10^5$ samples to evaluate $\fidelity_c$ in Eq.(\ref{eq:fc}) for all the simulations done in this work.
    }
    \label{fig:fig3}
\end{figure}
Now we investigate the efficiency of our method. We want to explore how the tomography complexity grows as the number of qubits $ N $ increases. Without losing generality, we consider $\rho$ as a mixed state representing the outcome of an unknown pure state $\rho'$ polluted by local depolarizing noise, characterized by the depolarizing strength $\gamma\in (0,1)$.  In particular, we focus on two cases: (1) $\rho'$ is a GHZ state, and (2) $\rho'$ is a W state. These two examples are also frequently used in literature to demonstrate the performance of other QST methods. As discussed earlier, given the unknown state $\rho$, the POVM $\{M^{\bold a}\}$ will generate a sample data $\{\bold a^{(j)} \}_{j=1}^{N_{\s}}$ with sample size $N_{\s}$. As long as $N_{\s}$ is sufficiently large, we are able to recover all information about $\rho$. Hence, the larger $N_s$ we choose, the larger $\mathcal{F}_c$ we can get. Then the efficiency of LM-QST can be measured by the minimum value of $N_{\s}^{\min}=N_{\s}^*$ such that the reconstructed $P(\bold{a})$ can achieve a fidelity $\fidelity_c\ge 0.99$. Since $N_{\s}^*$ is a function of the system size $N$ and the strength of depolarization noise $\gamma$, we plot $N_{\s}^*$ against $N$ and $\gamma$ to study the sample complexity of our method, as in Fig.~\ref{fig:fig3}. We find that in both cases and for different values of $\gamma$, $N_{\s}^*$ is approximately a linear function of $N$. In particular, to achieve $\fidelity_c \ge 0.99$, in both cases (1) and (2) with $N=50$, our method only requires a sample data with $N_{\s}=2\times 10^4$, compared to $N_{\s}=8 \times 10^4$ for the neural-network based approach in~\cite{CarrasquillaAolita2019}, demonstrating its great potential in sample complexity.

\textit{ LM-QST for a classical of similar states} -- A major benefit of our LM-QST from language modeling, distinguished from existing neural network-based QST methods, is that it can naturally accept integer sequences from similar quantum states with different sizes for training, and then generate samples with different lengths. Such a multi-state model QST works as follows.

\begin{table}
\caption{Distribution of the number of training data, $|\samples_n|$, from similar states with different system sizes $n$. We have considered two different types of quantum states: GHZ states and the ground states of transverse field Ising chain, and chosen their distribution against $n$ to be the same as the \textit{WikiText} dataset.}
\label{tab:distribution}
\begin{tabular}{@{}cccccc@{}}
\toprule
\multicolumn{2}{c}{WikiText}              & \multicolumn{2}{c}{GHZ}            & \multicolumn{2}{c}{TFIC ground states} \\ \midrule
length  & \multicolumn{1}{c|}{proportion} & length & \multicolumn{1}{c|}{proportion} & length      & proportion      \\ \midrule
2-50    & 0.41490                         & 2-5    & 0.41490                         & 2-5         & 0.41490         \\
51-100  & 0.15452                         & 6-10   & 0.15452                         & 6-10        & 0.15452         \\
101-150 & 0.18046                         & 11-15  & 0.18046                         & 11-15       & 0.18046         \\
151-200 & 0.12480                         & 16-20  & 0.12480                         & 16-20       & 0.12480         \\
201-250 & 0.06842                         & 21-25  & 0.06842                         & 21-25       & 0.06842         \\
251-300 & 0.03427                         & 26-30  & 0.03427                         & 26-30       & 0.03427         \\
301-350 & 0.01393                         & 31-35  & 0.01393                         & 31-35       & 0.01393         \\
351-400 & 0.00547                         & 35-40  & 0.00547                         & 35-40       & 0.00547         \\
401-450 & 0.00193                         & 41-45  & 0.00193                         & 41-45       & 0.00193         \\
451-500 & 0.00126                         & 46-50  & 0.00126                         & 46-50       & 0.00126         \\ \bottomrule
\end{tabular}
\end{table}

In the training stage, the model expects input sequences of the same predefined length $\Nmax$ as shown in Fig.~\ref{model-struct}. Given a sample $\bold{a}$, we first generate an input sequence $(\sos, a_1, \dots, a_N, \eos)$. After that, there are two commonly used approaches to enforce the same length for each input sequence~\cite{radford2019gpt2}: i) appending $\pad$s at the end of each sequence, namely
\begin{align}\label{eq:padding}
\bold{b} = (\sos, a_1, \dots, a_N, \eos, \pad, \dots);
\end{align}
or ii) concatenating all the input sequences into a single long sequence, and then chopping the long sequence into sequences of length $\Nmax$.
We use the latter approach since its encoding is more efficient compared to the former. Once the model is trained, the probability for any sample $\bold{a}$ with $N\leq\Nmax-2$ can be computed by evaluating $P(\bold{b})$ with $\bold{b}$ in Eq.~(\ref{eq:padding}). The classical fidelity for each model during training is estimated based on $10^5$ samples drawn from the trained model.(See Supplementary for more details)

\begin{figure}[htbp]
    \centering
    \includegraphics[width=\columnwidth]{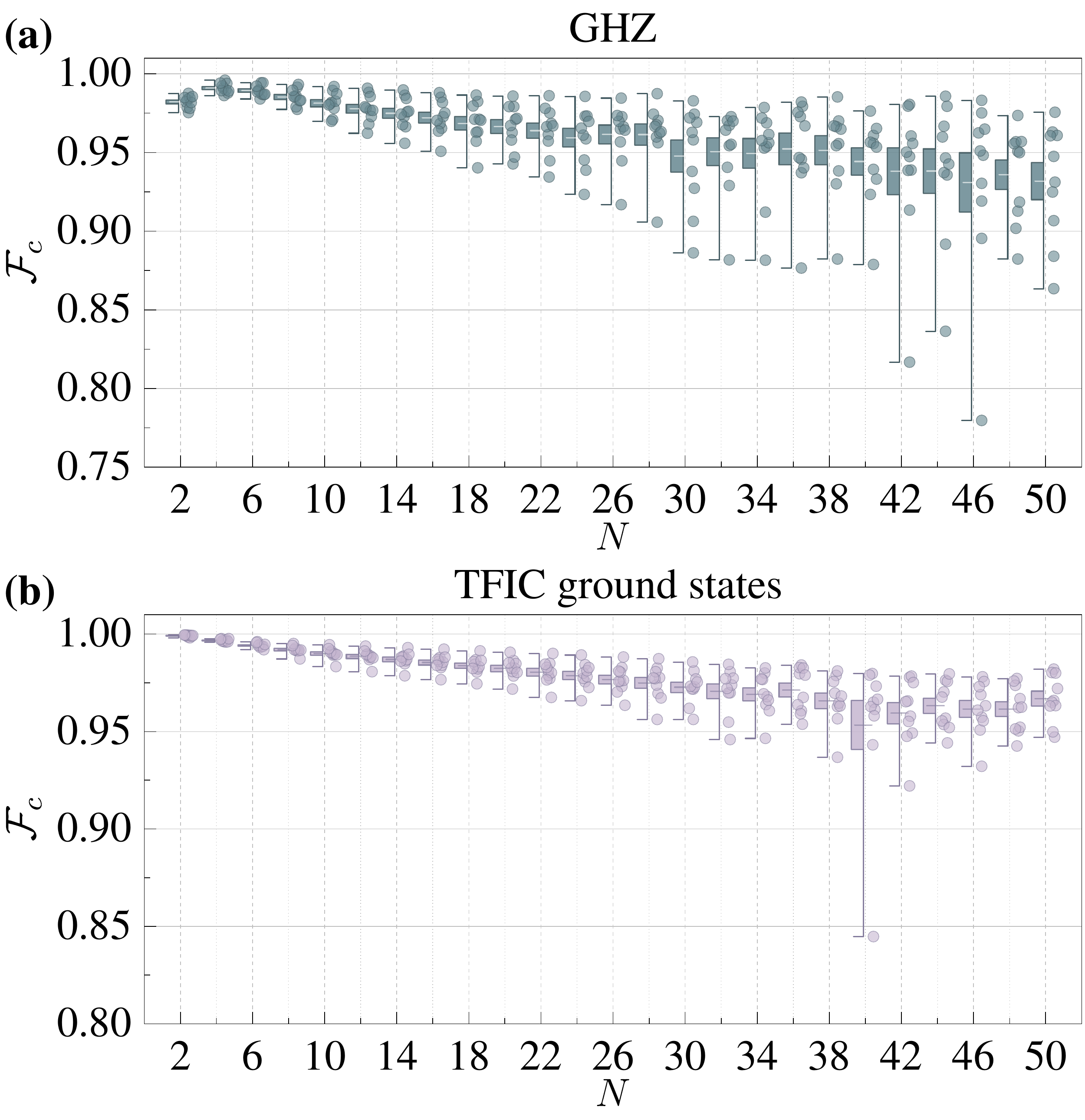}
    \caption{Boxplot showing the the classical fidelity $\fidelity_c$ as a function of the system size $N$ for (a) the GHZ states and (b) the ground states of the transverse field Ising chain. For each fixed number of qubit, there are $10$ circles which correspond to the classical fidelities computed by the $10$ models trained from independently generated synthetic training datasets. The box shows the standard derivation and the line in the middle of the box shows the mean, while the lines on the boundaries show the maximum and minimum. 
    % \gcc{All the models have been trained with the same training settings with $d=256$, $L=8$, $h=4$. Learning rate have been set to $10^{-3}$, we also used Adam optimizer.}
    We have used $L=8$ and $d=256$ in these simulations.
    }
    \label{fig:4}
\end{figure}

Now, we consider a family of states composed by (1) GHZ states with different $N$, and (2) the ground states of the transverse field Ising chain (TFIC) $\Hop=\sum_{i=1}^{N-1}Z_iZ_{i+1}+\sum_{i=1}^N{X_i}$, with different $N$.
First, we generate a hybrid training dataset from a class of similar states of different lengths, denoted as $\samples = \{ \samples_2, \samples_3, \dots, \samples_{50} \}$ (the subscripts denote the number of qubits of the corresponding quantum state), where the sizes $|\samples_n|$ follow the same distribution as the \textit{WikiText} dataset~\cite{merity2016wikitext} which is a collection of over $100$ million tokens extracted from the set of verified Good and Featured articles on Wikipedia, as shown in TABLE~\ref{tab:distribution}. The numerical results are shown as a box chart in Fig.~\ref{fig:4}, where we have plotted the results for $10$ models trained from independently generated synthetic training datasets (each with $10^5$ samples) to avoid any bias. 
Interestingly, from Fig.~\ref{fig:4} we can see that the reconstruction performs fairly well for both type of states, and for different number of qubits ranging from $2$ to $50$, even if there is only a tiny portion of training data for large systems. In particular, the average $\fidelity_c$ is above $92\%$ for GHZ states and above $95\%$ for TFIC ground states. In both cases, the average $\fidelity_c$ is high for small systems and decreases moderately for larger systems, which could be partially due to that our training data is concentrated on small systems. For larger systems the standard deviation around the mean values also becomes larger, but the majority of the classical fidelities for the $10$ models are still in the high fidelity regions except one or two exceptional points for which $\fidelity_c < 90\%$.

In summary, we have proposed a GPT-2-inspired transformer variation model for QST based on the insight that there is a close similarity between QST and the language modeling task. Using synthetic data for typical quantum states (GHZ and W states under local depolarization noises), we demonstrate that the number of samples required by our method to achieve high reconstruction accuracy (with classical fidelity $\fidelity_c> 0.99$) scales approximately linearly with the system sizes and that only $O(10^4)$ samples are required for all the quantum states considered (with up to $50$ qubits). As a prominent feature of our method benefiting from language modeling, we also demonstrate that we can reconstruct a classical of similar quantum states simultaneously using a hybrid training dataset mainly concentrated on smaller systems.

\begin{acknowledgments}
% L. Z. and X. W. are supported by the National Natural Science Foundation of China (Grant No.~92265208), and the National Key R\&D Program of China (Grant No.~2018YFA0306703). C. G. acknowledges support from National Natural Science Foundation of China under Grants No.~11805279, No.~12074117, No.~61833010 and No.~12061131011.
We acknowledge support from the National Natural Science Foundation of China under Grant No.~92265208, No.~11805279, No.~12074117, No.~61833010 and No.~12061131011, and the National Key R\&D Program of China under Grant No.~2018YFA0306703. 
\end{acknowledgments}

\bibliographystyle{apsrev4-1}
\bibliography{gcrefs}% Produces the bibliography via BibTeX.

\end{document}

\title{Supplementary Information: Quantum State Tomography as Language Modeling Task}

\date{\today}

\maketitle

\section{Constructing density matrix from the Probability distribution}
Here we first show the details of the conversion from the multi-variate probability distribution $P(\bold{a})$ to the density matrix $\rhoop$ for self-completeness of this work, which can also be found in Ref.~\cite{CarrasquillaAolita2019}. 
In this work we focus on informationally complete (IC) positive operator-valued measures (POVMs) instead of projective measurements, while the former can be systematically implemented as long as one can implement the latter experimentally, with the help of auxiliary qubits~\cite{NielsenChuang}. IC-POVM has the important property that every quantum state is uniquely determined by its measurement statistics, thus allowing an easier theoretical analysis.

We assume that the IC-POVM we choose contains $m$ operators denoted as $\{M^{(a)}\}$, with $a \in \{0,1,\dots, m-1\}$ labels the corresponding measurement outcomes. For an $N$-qubit quantum system, the possible measurement operators can be written in the tensor product form:
\begin{align}
M^{\bold{a}}=M^{a_1} \otimes M^{a_2} \dots \otimes M^{a_i} \dots \otimes M^{a_N},
\end{align}
where $a_i$ labels the local measurement outcome on qubit $i$.
Given an unknown $N$-qubit quantum state $\rhoop$, one can perform an IC-POVM locally on each qubit of it, and the probability that one obtains a sequence of measurement outcomes $\bold{a} = (a_1,a_2,\dots, a_N)$ can be computed as
% we prepare $N_s$ copies of $\ket{\psi}$. If we measure one copy of $\ket{\psi}$ with our prescribed measurement operator, the state will collapse to a certain outcome, we use sequence $\mathbf{a}=(a_1,a_2,\dots, a_i,\dots, a_N)$ to represent it, note $a_i \in \{0,1,2,..., m-1\}$ . By Born's rule, the probability of state $\ket{\psi}$ collapse to measurement outcome sequence $\mathbf{b}$ is
\begin{align} \label{eq1}
P(\bold{a})=\trace\left(\rhoop M^{\bold{a}}\right).
\end{align}
Note that all the IC-POVM $\{M^{\bold{a}}\}$ can span the entire space of the bounded-norm linear operators on the Hilbert space, therefore we can rewrite the density matrix $\rhoop$ as a linear combination of $M^{\bold{a}}$:
\begin{align}\label{eq2}
\rhoop = \sum_{\bold{a}}q_{\bold{a}} M^{\bold{a}},
\end{align} 
with the coefficients $q_{\bold{a}}$ to be determined.
Now we substitute Eq.(\ref{eq2}) into Eq.(\ref{eq1}) and get
\begin{align}\label{eq3}
P(\bold{a}') &= \trace\left(\sum_{\bold{a}}q_{\bold{a}} M^{\bold{a}}M^{\bold{a}'}\right) =\sum_{\bold{a}}  q_{\bold{a}}\trace \left(M^{\bold{a}}M^{\bold{a}'}\right).
 \end{align}
Denoting $T_{\bold{a},\bold{a}'}\equiv \trace\left(M^{\bold{a}}M^{\bold{a}'}\right)$, Eq.(\ref{eq3}) can be rewritten as:
\begin{align}\label{eq4}
 P(\bold{a}')&=\sum_{\bold{a}}q_{\bold{a}}T_{\bold{a},\bold{a}'}.
 \end{align}
Inverting $T_{\bold{a},\bold{a}'}$ (Moore–Penrose pseudo inverse is understood if $T_{\bold{a},\bold{a}'}$ is not explicitly invertible), we get
\begin{align}\label{eq5}
q_{\bold{a}} = \sum_{\bold{a}'}P(\bold{a}')T_{\bold{a},\bold{a}'}^{-1}
\end{align}
Substitute Eq.(\ref{eq5}) into Eq.(\ref{eq2}), we get
\begin{align}\label{eq6}
\rhoop = \sum_{\bold{a}}\sum_{\bold{a}'}P(\bold{a}')T_{\bold{a},\bold{a}'}^{-1}M^{\bold{a}}.
\end{align}
Therefore once the probability distribution $P(\bold{a}')$ is obtained, we can compute the density matrix $\rhoop$ using Eq.(\ref{eq6}) in principle.
% With IC-POVM operators $\{M^{(\mathbf{a})}\}$, the left part is to get the probabilities of all possible measurement outcome sequence $\{p(\mathbf{b})\}_{\mathbf{b}}$, then we can use (\ref{eq6}) to reconstruct the density matrix of corresponding quantum state $\ket{\psi}$.
% In this way, the task has become a probability estimation problem.

\section{Model details}
In the main text we have given a macroscopic description of the two building blocks, the embedding layer and the decoder block, used in our model. Here we will show the details of their structures for clearness.

\begin{figure}[ht]
    \centering
    \includegraphics[width=0.6\columnwidth]{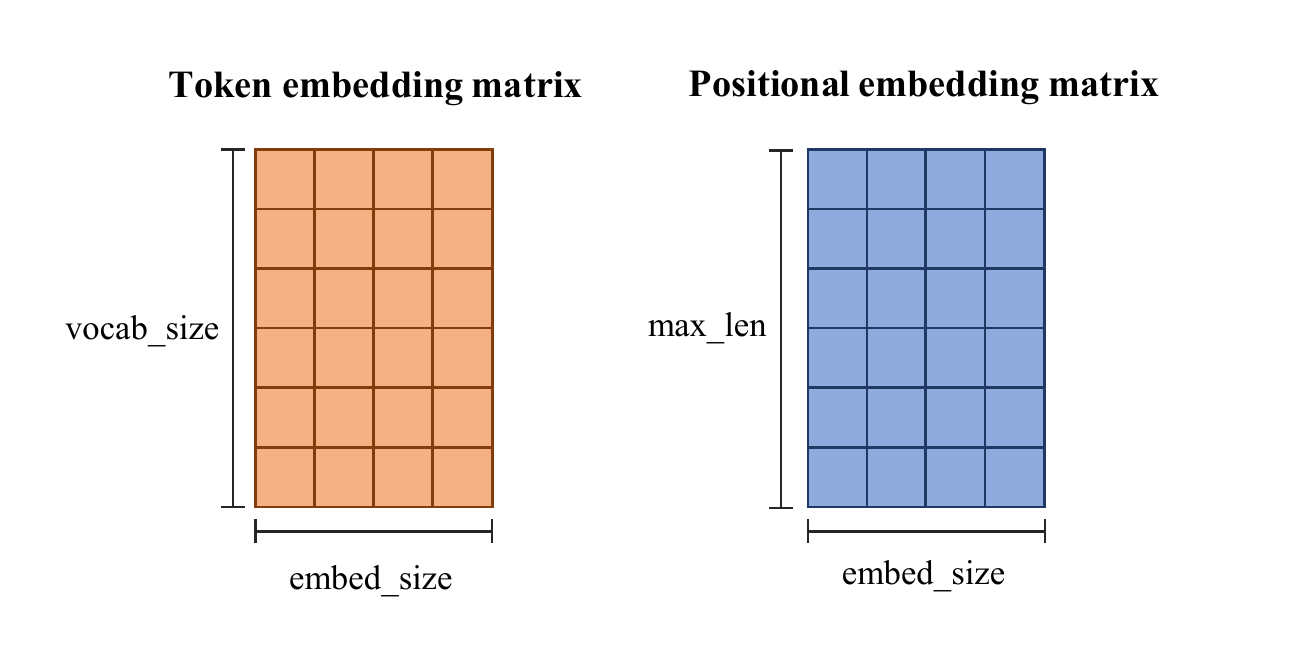}
    \caption{The token embedding matrix and the positional embedding matrix in the embedding layer, where vocab\_size denotes the vocabulary size, max\_len denotes the maximum input sequence length and embed\_size denotes the embedding size which is a hyperparameter in the embedding layer.
     }
\label{embedding_matrices}
\end{figure}

The embedding layer is parameterized by two matrices:  the token embedding matrix and the positional embedding matrix as shown in Fig.~\ref{embedding_matrices}. The token embedding matrix acts as a look-up dictionary with each row representing a word, the size of which is the vocabulary size (denoted as $|\voca|$ in the main text) times the embedding size denoted as $d$. Similar to the token embedding matrix, each row of the positional embedding matrix encodes the sequential order of the words, and the size of which is the maximum sequence length $\Nmax$ times the embedding size $d$. We have also use the function outlined in the original transformer paper to build the positional embedding matrix~\cite{2017attention}, which is treated as constant during the optimization. Given an input sequence $\bold{b}=(b_1,b_2,\dots, b_j,\dots)$, the embedding layer outputs a context vector $\bold{w}_j$ which is the $b_j$-th row of the token embedding matrix, and outputs a positional vector $\bold{p}_j$ which is the $j$-th row of the positional embedding matrix for each $1\leq j\leq \Nmax$.

\begin{figure}[ht]
    \centering
    \includegraphics[width=0.6\columnwidth]{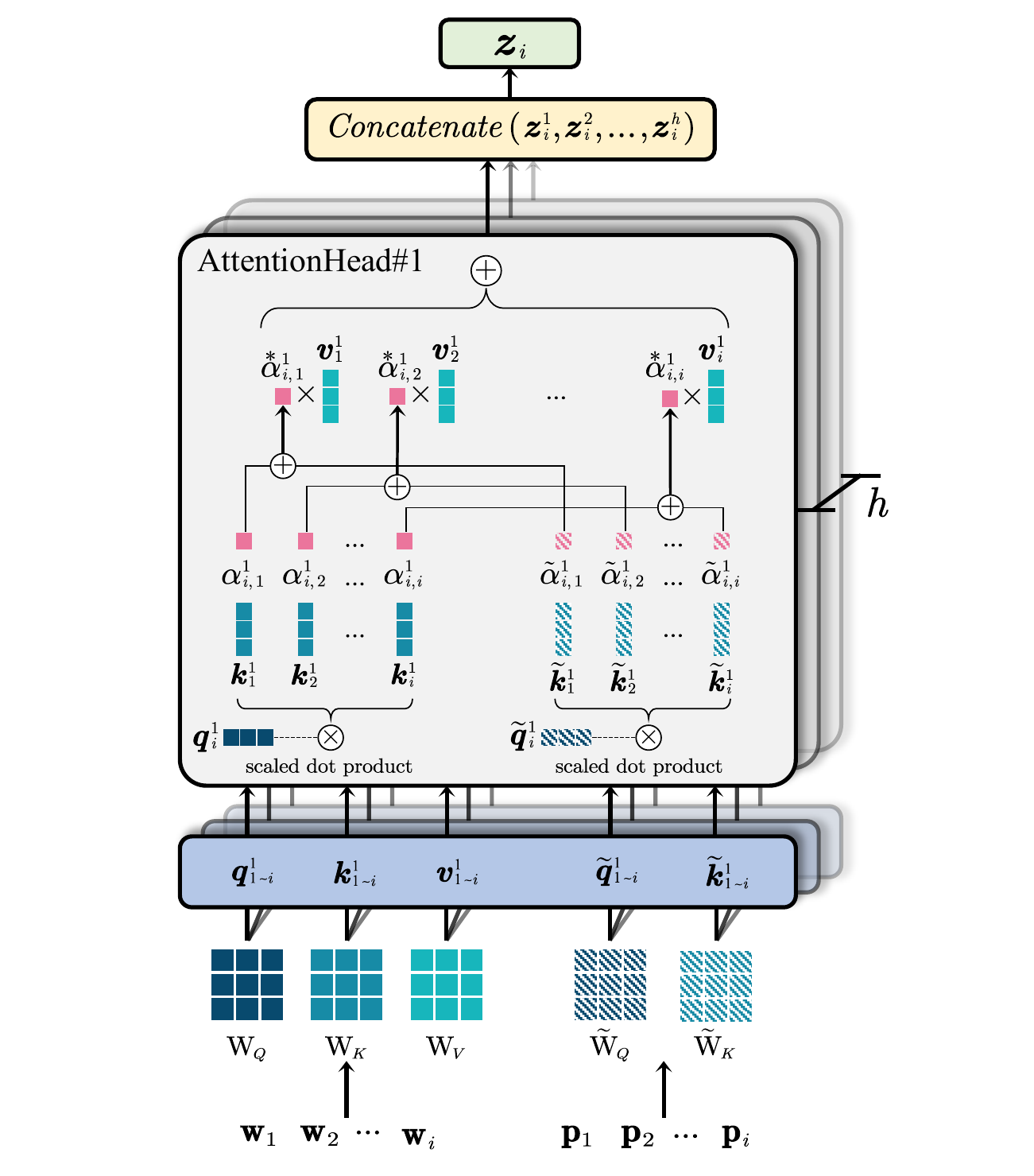}
    \caption{The computation flow in the masked multi-head self-attention layer to produce the $i$-th output context vector, denoted as $\bold{z}_i$, where all the input context vectors and positional vectors from $1$ to $i$, denoted as $\bold{w}_{1\sim i}$ and $\bold{p}_{1\sim i}$, are used as inputs.
     }
\label{attention-details}
\end{figure}

\begin{figure}[htbp]
    \centering
    \includegraphics[width=0.6\columnwidth]{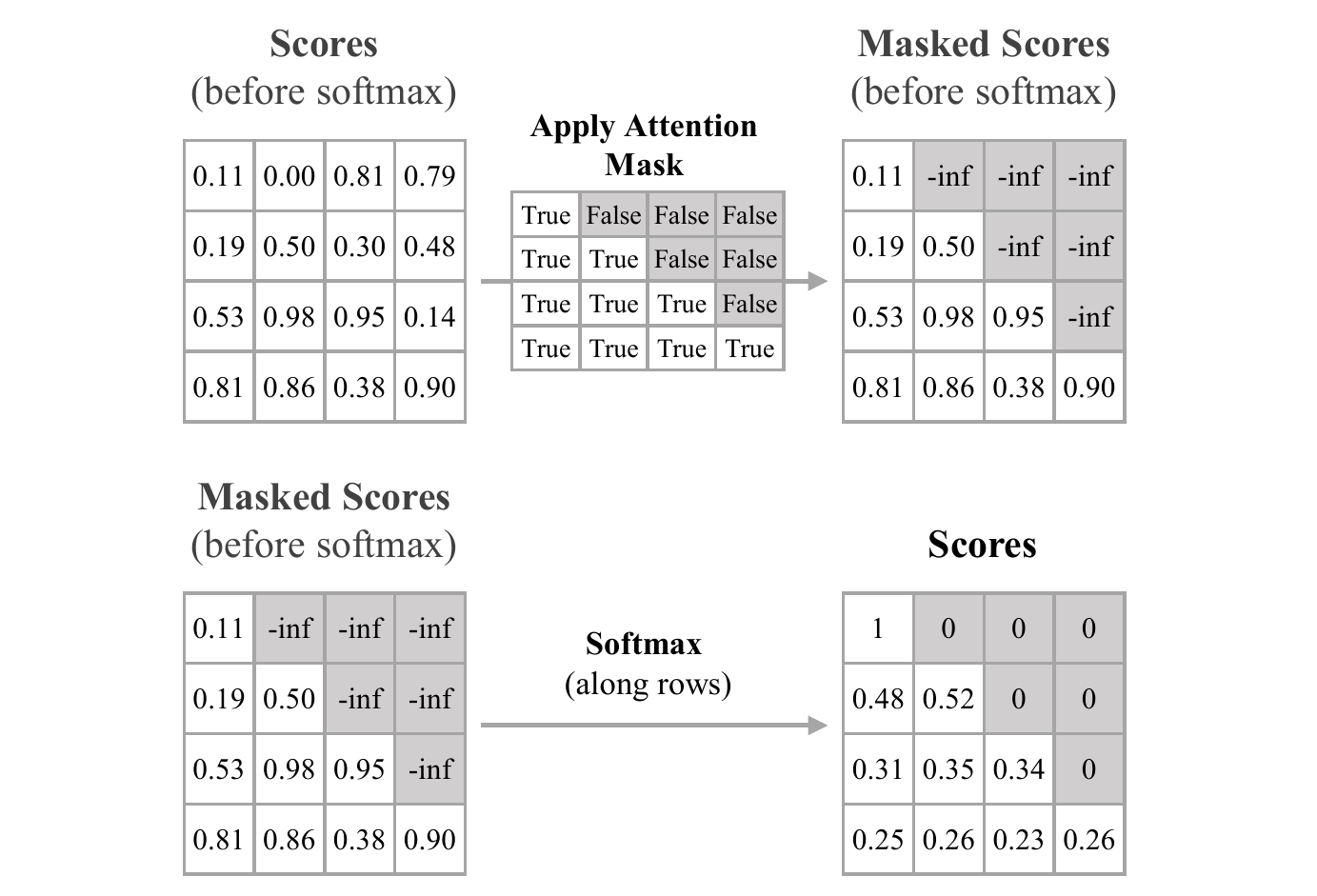}
    \caption{The attention mask matrix which eliminates the influences of future tokens to the current token by setting the upper triangular entries of the score matrix to zero.
     }
\label{masked-attention}
\end{figure}

The stacked decoder blocks are the backbone of our model. Each decoder block accepts a sequence of context vectors for last layer and the positional vectors as inputs, and outputs a new sequence of context vectors while keeping the positional vectors unchanged. Each decoder block consists of a masked multi-head self-attention layer, followed by a feed forward layer. 
% Let's now dive deeper into decoder blocks. The masked multi-head self-attention layer is the most crucial component of the decoder block. (Note that this mask has a different meaning than that in BERT\gcc{[BERT]}.) One significant advantage of using the self-attention layer is that it enables the model to focus on the relevant part of an input sequence as needed. By scoring the relevance between words and summing their vector representations, the self-attention layer can incorporate the model's understanding of associated parts before processing a particular word, thus explaining the context of that word. 
The masked multi-head self-attention layer is the most crucial component of the decoder block.
% In most cases, the position vectors are directly added to the context vectors at the embedding layers and are not subsequently used. However, we consider that the individual qubits in a quantum state are extremely susceptible to the position. 
% For example, a two-qubit measurement operator $M$ is the tensor product of two single-qubit measurement operators, $M = M_1\otimes M_2$. Changing the order of these two different operators will most likely have a completely different effect on a quantum state. 
Overall, it outputs a sequences of context vectors, where the $i$-th output context vector depends on all the input context vectors and positional vectors from $1$ to $i$, denoted as $\bold{w}_{1\sim i}$ and $\bold{p}_{1\sim i}$ respectively, as shown in Fig.~\ref{attention-details}. 
In literatures\cite{2017attention,devlin2018bert, radford2018gpt, radford2019gpt2}, the positional vectors are often added to the context vectors at the embedding layers and will not be directly used afterwards. However, in the context of QST, the quantum state could be extremely sensitive to the positional relations among individual qubits. For example, considering a two-qubit measurement operator $M^{\bold{a}} = M^{a_1} \otimes M^{a_2}$ on the quantum state $\rhoop$, changing the order of $M^{a_1}$ and $M^{a_2}$ will likely have a completely different effect on $\rhoop$. 
Therefore we treat context vectors and positional vectors separately in every decoder block to avoid information confusion.
% For this concern, we decide to treat context vectors and position vectors as two different types of inputs in every decoder block to avoid information confusion. fig.\ref{attention-details} illustrates the detailed attention layer we use in our decoder block. Suppose the attention layer of first decoder block is now handling token $a_i$ of sequence $a$, then the inputs are the context vectors $\boldsymbol{w}_{1\sim N}$ and the position vectors $\boldsymbol{p}_{1\sim N}$ generated by the previous embedding layers. 
% Notably, we are processing these two different classes of vectors separately. 
Concretely, the context vectors are multiplied with three projection matrices to get the query, key, and value vectors:
\begin{align}
\boldsymbol{q}_i &= \bold{w}_i W_Q ; \\
\boldsymbol{k}_i &= \bold{w}_i W_K ; \\
\boldsymbol{v}_i &= \bold{w}_i W_V . 
\end{align}
Similarly, the positional vectors are multiplied with two different projection matrices to get the positional query, key vectors:
\begin{align}
\tilde{\boldsymbol{q}}_i = \bold{p}_i \tilde{W}_Q ;  \\
\tilde{\boldsymbol{k}}_i = \bold{p}_i \tilde{W}_K .  
\end{align}
Then we uniformly split each projected vector into $h$ segments (we have chosen $h=4$ in all the simulations done in this work). These segments will be fed into $h$ heads in the attention layer separately. This is also the reason why this layer is referred to as multi-head. As shown in Fig.~\ref{attention-details}, the first attention head is responsible for processing the first segment of all the vectors $\boldsymbol{q}_{1\sim i}$, $\boldsymbol{k}_{1\sim i}$, $\boldsymbol{v}_{1\sim i}$, $\tilde{\boldsymbol{q}}_{1\sim i}$ and $\tilde{\boldsymbol{k}}_{1\sim i}$, which are denoted as $\boldsymbol{q}_{1\sim i}^{1}$, $\boldsymbol{k}_{1\sim i}^{1}$, $\boldsymbol{v}_{1\sim i}^{1}$, $\tilde{\boldsymbol{q}}_{1\sim i}^{1}$ and $\tilde{\boldsymbol{k}}_{1\sim i}^{1}$ respectively. In the first attention head (and similarly for the rest heads) two scores are used to evaluate how relevant each token is to the token $a_i$, namely the context scores $\alpha_{i,j}^{1} $ and the positional scores $\tilde{\alpha}_{i,j}^{1}$ which are computed as:
\begin{align}
\alpha_{i,j}^{1} &= \frac{1}{\sqrt{2d}}\boldsymbol{q}_i^1 (\boldsymbol{k}_j^{1 })^T; \\ 
\tilde{\alpha}_{i,j}^{1}&= \frac{1}{\sqrt{2d}}\tilde{\boldsymbol{q}}_i^1(\tilde{\boldsymbol{k}}_j^{1 })^T,
 \end{align}
where $j\in \{1,2,\dots, i\}$, $\bold{x}^T$ means the transpose of the vector $\bold{x}$, and $1/\sqrt{2d}$ is a scaling factor. The two scores are added together to get the score
\begin{align}\label{eq:score}
\accentset{\ast}{\alpha}_{i,j}^{1}=\alpha_{i,j}^{1}+\tilde{\alpha}_{i,j}^{1}.
\end{align}
The term ``masked'' refers to the fact that the model always scores the effects of the future tokens on the current token as zero. This masking is usually implemented as a matrix known as the attention mask, which is demonstrated in Fig.~\ref{masked-attention}. 
After we obtain the score in Eq.(\ref{eq:score}), we apply the attention mask on it to eliminate the unwanted influences from the future.
After that, we apply ${\rm softmax}$ on each row to yield the final scores.  
Finally, we multiply each token's value vectors by its final scores and sum them together to get the output vector
\begin{align}
\boldsymbol{z}_i^1 = \sum_{j=1}^{j=i}\accentset{\ast}{\alpha}_{i,j}^{1} \boldsymbol{v}_j
\end{align}
of the first attention head. The output vectors from all the heads are concatenated together to generate the $i$-th output context vector
\begin{align}
\boldsymbol{z}_i=Concat\{\boldsymbol{z}_i^1, \boldsymbol{z}_i^2,\dots,\boldsymbol{z}_i^h\}.
\end{align}

After the masked multi-head self-attention layer,  the output context vectors will go through a feed-forward layer before being sent into the next layer. The feed-forward network consists of two fully-connected layers, where the first layer increases the sizes of the context vectors from $d$ to $4d$, and the second one decreases them from $4d$ to $d$. This feed-forward layer is the same as the original one used in the transformer decoder~\cite{2017attention}. The main reason to use a feed-forward layer is to make the context vectors more expressive through nonlinear transformations, since what we do in the masked multi-head self-attention layer is mainly matrix multiplication (and scaled dot-product). The feed-forward layer can be written as:
\begin{align}
   \textrm{FFN}(\boldsymbol{z}_i)=\relu(\boldsymbol{z}_iW_1+b_1)W_2 + b_2 
\end{align}
where $\relu$ represents the ReLU activation function, $W_1$ and $W_2$ are the two matrices used in feed-forward layer, $b_1$ and $b_2$ are the biases. 

After all the decoder blocks have been applied, there is a second embedding layer which is the transpose of the first embedding layer, which transforms each input context vector of size $d$ into the final output vector of size $|\voca|$ again. The $i$-th final output vector is interpreted as the conditional probability distribution of the $(i+1)$-th token on all the previous tokens, denoted as $p(x_{i+1}|b_1, \dots, b_i)$. As a result, the total number of tunable parameters in a single decoder block is $13d^2 + 5d $, and the total number of parameters in our model is $(13d^2 + 5d)L + d|\voca|$.

\section{Exact sampling from trained model}
Our model satisfies an important property of being \textit{auto-regressive}, which is because that by construction the current token will only be affected by the previous tokens due to the attention mask in Fig.~\ref{masked-attention}. In the training stage, after all the conditional probability distributions from $p(x_2|b_1)$ to $p(x_{\Nmax+1}|b_1, \dots, b_S)$ have been obtained from the last embedding layer, we compute the probability of the input sequence by selecting the  $b_{i+1}$-th element of $p(x_{i+1}|b_1, \dots, b_{i})$ for all $1\leq i\leq \Nmax-1$ and multiply them together (the last output is neglected):
\begin{align}
P(\bold{b}) = p(b_2|b_1) p(b_3|b_1, b_2), \dots, p(b_{\Nmax}|b_1, \dots, b_{\Nmax-1}). 
\end{align}

\begin{figure}[htbp]
    \centering
    \includegraphics[width=0.6\columnwidth]{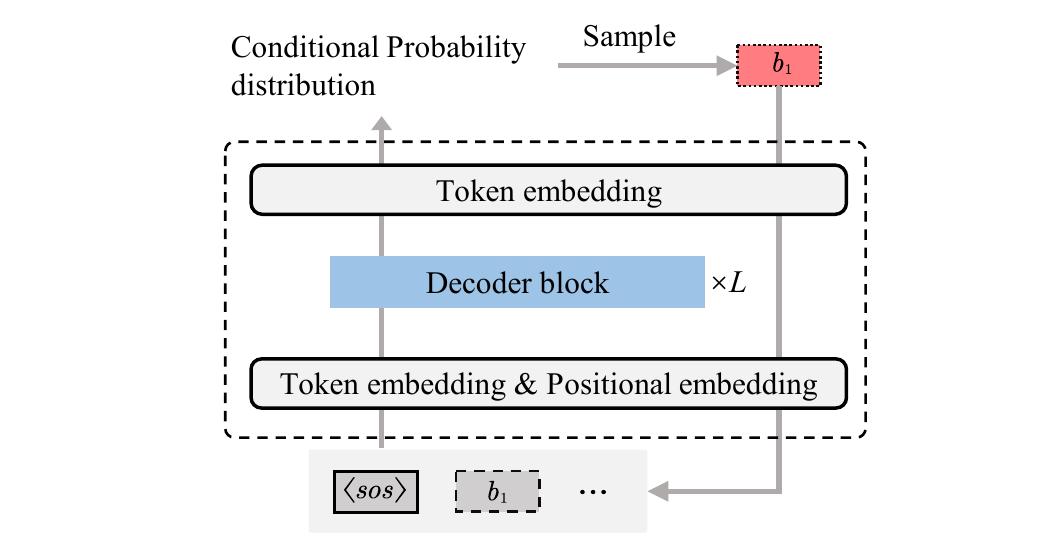}
    \caption{Exact sampling algorithm with a trained model, which can generate any sample with length $N < \Nmax$.
     }
\label{sample-process}
\end{figure}
Once the model has been trained, one can generate samples with lengths smaller than $\Nmax$ as follows. One first feeds the sequence $\bold{b} = (\sos, \pad, \dots, \pad)$ of length $\Nmax$ into our model, then the first output of our model is the conditional probability distribution of the second token on $\sos$, namely $p(x_2|\sos)$. After that, we perform a local sampling on $p(x_2|\sos)$ restricted in the subspace spanned by all the local measurement outcomes, that is, the subspace $(0, 1, \dots, m-1)$. This local sampling can be done exactly since $m=4$ in our case (In comparison in NLP $m$ could be tens of thousands and one may need to resort to approximate methods), which produces the second token $b_2$. Similarly, one can feed $\bold{b} = (\sos, b_2, \pad, \dots, \pad)$ into our model again and obtain the next token $b_3$. Repeating this procedure, one can generate any sequence $\bold{b} = (\sos, b_2, b_3, \dots, b_{N+1})$ for any $N < \Nmax$, from which we get a sample $\bold{a} = (b_2, b_3, \dots, b_{N+1})$. This exact sampling algorithm is demonstrated in Fig.~\ref{sample-process}, as a standard practice for language models.

\section{Convergence analysis for small-scale GHZ and W states}

In the main text we have mostly focused on large systems and only considered the classical fidelity $\fidelity_c$ as the reconstruction quality measure. Here we consider smaller-scale GHZ and W states, and study the convergence of our algorithm, as well as two other measures for the reconstruction quality: 1) the Kullback-Leibler(KL) divergence defined as
\begin{align}
D_{KL}(P||P_{\true}) = \mathbb{E}_{\bold{a}\sim P_{\true}}\left[\log\left(\frac{P(\bold{a})}{P_{\true}(\bold{a})} \right) \right],
\end{align}
which measures how much the reconstructed distribution $P(\bold{a})$ diverges from the target distribution $P_{\true}(\bold{a})$; and 2) the Perplexity (PPL) which is one of the most typical metrics for evaluating language models~\cite{jelinek1977perplexity}, defined as
\begin{align}
\PPL(\bold{a}) = e^{-\frac{1}{N} \sum_{i=1}^N \log(p(a_{i+1}|a_1, \dots, a_i)) }.
\end{align}

\begin{figure}[htbp]
    \centering
    \includegraphics[width=0.9\columnwidth]{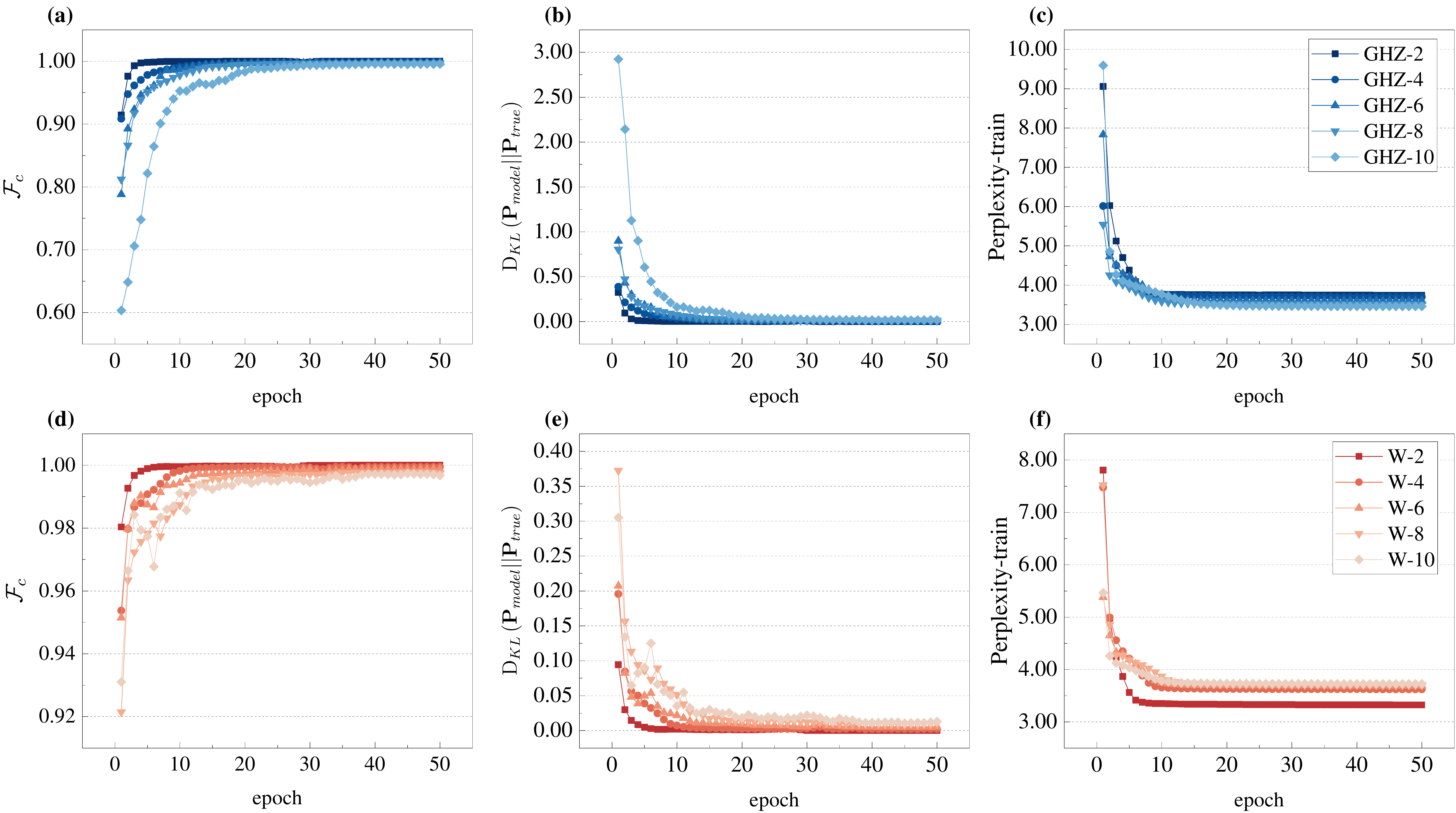}
    \caption{(a,b,c) Convergence of the classical fidelity $\fidelity_c$, the KL divergence $D_{KL}$, and the Perplexity $\PPL$ as a function of the number of epochs for the GHZ states with $N=2,4,6,8,10$ respectively. (d,e,f) Similar to (a,b,c) for the W states. 
    % \gcc{All the models have been trained with $N_s = 2\times 10^4$ samples. We have also used $d=64$, $L=4$, $h=4$.}
    All the models in these simulations have been trained with $N_s = 2\times 10^4$ samples, for which we have also used $L=4$ and $d=64$. 
     }
\label{ghz-pure}
\end{figure}

We study the convergences of the three quantities, $\fidelity_c$, $D_{KL}$ and $\PPL$ against the number of epochs during the iterative optimization for pure GHZ states and W states with $N=2,4,6,8,10$ respectively, and the results are shown in Fig.\ref{ghz-pure}. We can see that all the reconstruction quality measures have well converged within $20$ epochs. 
% Here we focus on small-scale GHZ and W states and study the detailed convergence of our algorithm, in terms of the classical fidelity used in the main text, as well as two other improtant measures.

\section{Effects of different noise models}

In the main text we have shown the classical fidelity $\fidelity_c$ for large-scale GHZ and W states under the local depolarization noise, with quantum channel
\begin{align}
\mathcal{E}^{dp} ( \rhoop ) = \left(1-p_1\right)\rhoop + \frac{p_1}{3}\left(  X\rhoop X + Y\rhoop Y + Z\rhoop Z\right ),
\end{align}
where $X$, $Y$ and $Z$ are the Pauli matrices and $p_1$ is the depolarization noise strength.
It could also be interesting to look at the performance of our LM-QST when applied to mixed quantum states under other types of noise models. Here we also consider the local bit flip noise with quantum channel
\begin{align}
\mathcal{E}^{bf}(\rhoop) = (1-p_2) \rhoop + p_2 X \rhoop X,
\end{align}
where $p_2$ is the noise strength. For completeness, we also consider the quantum fidelity $\fidelity_q$ defined between two quantum states $\tau$ and $\sigma$:
\begin{align}\label{eq:quantumfidelity}
\fidelity_q(\tau, \sigma) = \left[\trace\left(\sqrt{\sqrt{\tau} \sigma \sqrt{\tau}}\right)\right]^2,
\end{align}
as a reconstruction quality measure for small quantum systems.
To compute Eq.(\ref{eq:quantumfidelity}), we will first convert the reconstructed probability distribution $P(\bold{a})$ into a density matrix $\tilde{\rhoop}$ using Eq.(\ref{eq6}), and then compute the quantum fidelity between $\tilde{\rhoop}$ and the target quantum state $\rhoop$.

\begin{figure}[htbp]
    \centering
    \includegraphics[width=0.8\columnwidth]{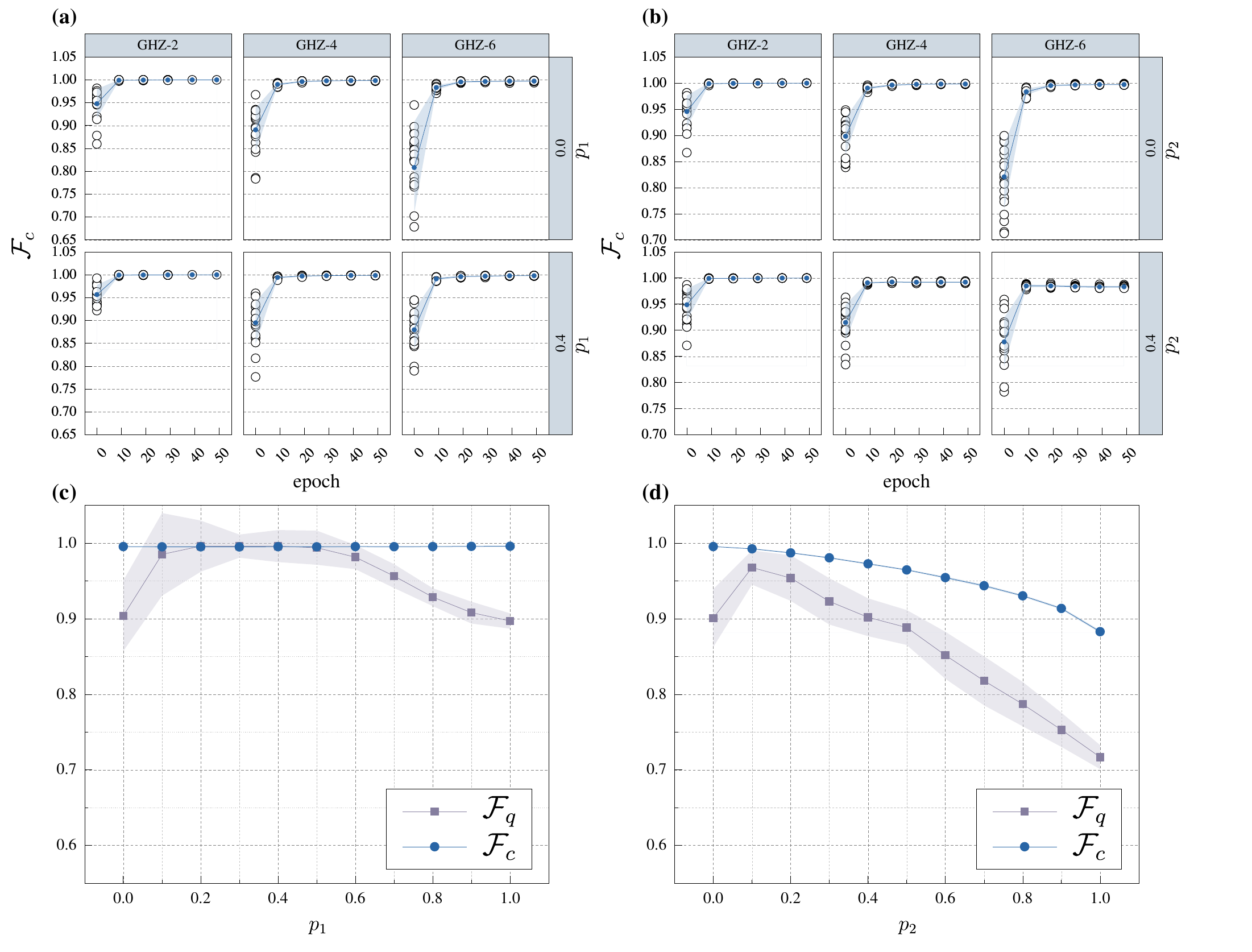}
    \caption{(a, b) Classical fidelity as a function of the number of epoch for the GHZ states under the depolarization noise in (a) and under the bit flip noise in (b). (c, d) The optimal classical and quantum fidelities as functions of the noise strength for (c) the depolarization noise and (d) the bit flip noise. In these simulations we have used $N_s = 2\times 10^4$, $L=4$ and $d=64$.
     }
\label{ghz-noisy}
\end{figure}

Concretely, we focus on the GHZ states under both depolarization and bit flip noises with $N=2,4,6$, and the results are shown in Fig.~\ref{ghz-noisy}. From Fig.~\ref{ghz-noisy}(a,b), we can see that both the classical and quantum fidelities converge to the optimal values with a similar trend, while in general the quantum fidelity converges slower and may converge to a lower value compared to the classical fidelity. In Fig.~\ref{ghz-noisy}(c,d) we show the final classical and quantum fidelities as a function of the noise strengths for $N=6$, from which we can see that for the bit flip noise both the classical and quantum fidelity are significantly lower than for the depolarization noise. While both of them still follow a similar trend as $f$ increases.

\section{Generation of the synthetic training datasets}
To generate training datasets for large pure quantum states, such as GHZ and W states with $50$ qubits, we first rewrite them as matrix product states (MPSs) and generate the measurement outcomes by sampling these MPSs (MPS also allows exact sampling~\cite{HanZhang2018}). We have used the ITensor library~\cite{itensor} to generate the MPS representations for the GHZ and W states, as well as to compute the ground states of the transverse field Ising chain as MPSs (where the density matrix renormalization group algorithm is used).
For pure quantum states under different noises, we have used the strategy in Ref.~\cite{TorlaiCarleo2018}.

% The ITensor library~\cite{itensor} was used to perform these simulations. As for the ground states of these spin Halmitonians states as MPOs, we performed the density matrix renormalization group(DMRG) algorithm to get their ground states(also in the form of MPSs), these processes are all implemented within the ITensor package. Once we obtain the ground states we can apply measurement operators on them to get the measurement outcome sequences. For noise test, we used the generation strategy adapted in \cite{TorlaiCarleo2018}. One can find our detailed codes for the generation of training datasets in \cite{sourcecode}.

\bibliographystyle{apsrev4-1}
\bibliography{gcrefs}